\documentclass[prl,twocolumn,floats,showpacs,superscriptaddress]{revtex4}
\usepackage{amssymb}
\usepackage{graphicx}
\usepackage{dcolumn}
\usepackage{amsmath}
\usepackage{bm}
\usepackage{epsfig}

\def\t{\theta}

\begin{document}

%%\title{$k$-core percolation and $k$-core organization of complex networks}
\title{Phase diagram of the three states Potts model with next nearest neighbor
interactions on the Bethe lattice}
\author{Nasir Ganikhodjaev}
\affiliation{Faculty of Science, IIUM, 25200, Kuantan,  Malaysia}
\author{Farrukh Mukhamedov}
\affiliation{Faculty of Science, IIUM, 25200, Kuantan, Malaysia}
\author{Chin Hee Pah}
\affiliation{Faculty of Science, IIUM, 25200, Kuantan, Malaysia}

\begin{abstract}

We have  found an exact phase diagram of the Potts model with next nearest neighbor
interactions on the Bethe lattice of order two. The diagram consists of five phases: ferromagnetic, paramagnetic, modulated,  antiphase and paramodulated, all meeting at the Lifshitz point i.e. $p=1/3$.
We report on a new phase which we denote as paramodulated, found at low temperatures and characterized by 2-periodic points of an one dimensional dynamical system lying inside the modulated phase. Such a phase, inherent in the Potts model has no analogues in the Ising setting.

\end{abstract}

%\pacs{05.10.-a, 05.40.-a, 05.50.+q, 87.18.Sn}
\maketitle

The $q$-state Potts model is one of the most studied models in
statistical mechanics due to its wide theoretical interest and
practical applications \cite{NS,Ba,DGM,Mo,T}.  The Potts
model \cite{Po} was introduced as a generalization of the Ising
model to more than two components and presently
encompasses a number of problems in statistical physics (see, e.g.
\cite{W}). The model is  structured richly enough to illustrate
almost every conceivable nuance of the subject. While some exact results
regarding certain properties of the model are known, most of them
are based on approximation methods. In fact, to date, no analytical solutions
on standard lattices are known to exist. Interestingly enough however, investigations on
phase transitions of spin models on hierarchical lattices provides
exact calculations of various physical quantities \cite{DGM,Mo,T,CMC}. Such studies on the hierarchical
lattices begun with development of the Migdal-Kadanoff
renormalization group method where the lattices emerged as
approximants of the ordinary crystal ones. It is believed that
several among its interesting thermal properties could persist for
regular lattices, for which the exact calculation is by far
intractable. In \cite{PLM,g}, the phase diagrams of the $q$-state
Potts models on the Bethe lattices were studied and the pure
phases of the the ferromagnetic Potts model were found.  The Bethe
lattices were fruitfully used, providing a deeper insight into the
behavior of the Potts models.

On the other hand, compared to the Ising models with competing
interactions \cite{Ising}, the Potts models with such interactions
on regular and trees are less studied \cite{NS,KF,PT,Ma}. In
\cite{Ta}, the phase diagram for the $q$-state Potts model is
constructed by means of the low-temperature expansion technique.
An infinite set of phases appears with the bifurcating structure
resembling the complete Devil's staircase. In \cite{BW,I} the
three-state Potts model with antiferromagnetic nearest-neighbor
and ferromagnetic next-nearest-neighbor interaction was
investigated within a mean-field theory.

To the best knowledge of the authors, $q$-state Potts model
with competing interactions on the Bethe lattices are not
well studied \cite{one}. In this Letter we present
a phase diagram of the three-state Potts model with next nearest
neighbor interactions on a Bethe lattice of order two. The similarity of results obtained for models defined on Bethe
lattices and on crystal lattices presents a strong motivation for the
study of models on Bethe lattices, since statistical mechanics on
such lattices presets many simplifying aspects that are absent in
models defined on crystal lattices. One of the useful ways to investigate models defined on trees  is to formulate them as dissipative mapping problems which allows us to use the techniques of the theory of dynamical systems. For the models
defined on crystal lattices, such a method does not lend itself to
a simpler solution of the problem since the metastable
configurations correspond to unstable orbits of the mapping
\cite{Au}.

%\section{Model}

Recall that the Bethe lattice $\Gamma^k$ of order $ k\geq 1 $ is
an infinite tree, i.e., a graph without cycles with exactly $ k+1 $ edges issuing from
each vertex. Let
$\Gamma^k=(V, \Lambda),$ where $V$ is the set of vertices of $
\Gamma^k$, $\Lambda$ is the set of edges of $ \Gamma^k$. Two
vertices $x$ and $y$ are called {\it nearest neighbors} if there
exists an edge $l\in\Lambda$ connecting them, which is denoted by
$l=<x,y>$. The distance $d(x,y), x,y\in V$, on the Bethe lattice,
is the number of edges in the shortest path from $x$ to $y$. For a
fixed $x^0\in V$ we set $V_n=\{x\in V| d(x,x^0)\leq n\}$ and $L_n$
denotes the set of edges in $V_n$. For the sake of simplicity we
put $|x|=d(x,x^0)$, $x\in V$. Two vertices $x,y\in V$ are called
{\it the second neighbors} if $d(x,y)=2$. The second neighbor
vertices $x$ and $y$ are called {\it next nearest neighbor} if
$|x|\neq |y|$ and is denoted by $\widetilde{>x,y<}$.

In this letter, we will consider a semi-infinite Bethe lattice
$\Gamma^2_+$ of order 2, i.e. an infinite graph without cycles
with 3 edges issuing from each vertex except for $x^0$ which has
only 2 edges. Considering the three-state Potts model with spin
values in $\Phi=\{1,2,3\}$, the relevant Hamiltonian with next nearest neighbor
 interactions has the form
\begin{equation}\label{ham}
H(\sigma)=-J_p\sum\limits_{
\widetilde{>x,y<}}\delta_{\sigma(x)\sigma(y)} -J_1
\sum\limits_{<x,y>}\delta_{\sigma(x)\sigma(y)}\end{equation} where
$J_p,J_1\in {R}$ are coupling constants and $\delta$ is the
Kronecker symbol. In what follows, we consider the case where $J_1>0$ and
$J_p<0$.

In order to produce the recurrent equations, we consider the relation of the
partition function on $V_n$ to the partition function on subsets
of $V_{n-1}$. Given the initial conditions on $V_1$, the recurrence equations indicate how their influence propagates down the tree. Let $Z^{(n)}(i_1,i_0,i_2)$ be the
partition function on $V_n$ where the spin in the root $x^0$ is
$i_0$ and the two spins in the proceeding ones are $i_1$ and
$i_2$, respectively. There are  27 different partition
functions $Z^{(n)}(i_1,i_0,i_2)$ and the partition function
$Z^{(n)}$ in volume $V_n$ can the be written as follows
\begin{equation*}
Z^{(n)}=\sum^{}_{i_1,i_0,i_2}Z^{(n)}(i_1,i_0,i_2).\end{equation*}

As shown in \cite{gmm} one can select only five independent
variables $Z^{(n)}(1,1,1)$, $Z^{(n)}(2,1,2)$, $Z^{(n)}(1,2,1)$,
$Z^{(n)}(2,2,2)$, $Z^{(n)}(3,2,3)$ and with the introduction of
new variables
\begin{eqnarray*}
u_1^{(n)}=\sqrt{Z^{(n)}(1,1,1)}, \
u_2^{(n)}=\sqrt{Z^{(n)}(2,1,2)}, \\
\ u_3^{(n)}=\sqrt{Z^{(n)}(1,2,1)}, \
u_4^{(n)}=\sqrt{Z^{(n)}(2,2,2)}, \\
u_5^{(n)}=\sqrt{Z^{(n)}(3,2,3)},
\end{eqnarray*}
straightforward calculations (see more detail \cite{gmm}) show
that one has
\begin{equation}\label{ZN}
Z^{(n)}=(u^{(n)}_1+2u^{(n)}_2)^2+
2(u^{(n)}_3+u^{(n)}_4+u^{(n)}_5)^2
\end{equation}
and
\begin{equation}\label{rec1}
\left\{
\begin{array}{lll}
u_1^{(n+1)}&=&\theta_1 (\theta_p u^{(n)}_1+2 u^{(n)}_2)^2, \\[2mm]
u_2^{(n+1)}&=& (\theta_p u^{(n)}_3+ u^{(n)}_4+u^{(n)}_5)^2,
\\[2mm]
u_3^{(n+1)}&=& ( u^{(n)}_1+(\theta_p+1) u^{(n)}_2)^2, \\[2mm]
u_4^{(n+1)}&=& \theta_1( u^{(n)}_3+ \theta_p u^{(n)}_4+
u^{(n)}_5)^2,
\\[2mm]
u_5^{(n+1)}&=& ( u^{(n)}_3+  u^{(n)}_4+ \theta_p u^{(n)}_5)^2.
\\[2mm]
\end{array}
\right.
\end{equation}
where $\theta_p=\exp(\beta J_p),\theta_1=\exp(\beta J_1).$

We rewrite \eqref{rec1} in the reduced variables
\begin{eqnarray*}
\begin{array}{lll}
x=\dfrac{2u_2+u_3+u_5}{u_1+u_4}, && y_1=\dfrac{u_1-u_4}{u_1+u_4},\\
y_2=\dfrac{u_2-u_3}{u_1+u_4}, &&
y_3=\dfrac{u_2-u_5}{u_1+u_4}
\end{array}
\end{eqnarray*}as follows
\small
\begin{equation}
\label{rc}\left\{
\begin{array}{rcl}
x^{(n+1)}&=&\dfrac{1}{2\theta_1D^{(n)}}[P(y^{(n)}_1,y^{(n)}_2,y^{(n)}_3)\\[3mm]
&&+((\theta_p+1)x^{(n)}+2-y^{(n)}_1- \theta_p
y^{(n)}_2\\&&-y^{(n)}_3)^2];
\\[3mm]
y^{(n+1)}_1&=&\dfrac{2}{D^{(n)}}(\theta_p+x^{(n)})(\theta_p
y^{(n)}_1+y^{(n)}_2+y^{(n)}_3);
\\[3mm]
y^{(n+1)}_2&=&-\dfrac{1}{\theta_1D^{(n)}}[y^{(n)}_1+\theta_p
y^{(n)}_2-y^{(n)}_3]\\[2mm]
&&\times
[2+(\theta_p+1)x^{(n)}\\&&-(\theta_p-1)(y^{(n)}_2-y^{(n)}_3)];
\\[3mm]
y^{(n+1)}_3&=&\dfrac{1}{\theta_1D^{(n)}}(\theta_p-1)(y^{(n)}_3-y^{(n)}_2)\\
&&\times[2+
(\theta_p+1)x^{(n)}-2y^{(n)}_1\\
&&-(\theta_p+1)(y^{(n)}_2+y^{(n)}_3)];
\end{array}
\right.
\end{equation}
\normalsize where
\begin{eqnarray*}
D^{(n)}&=&(\theta_p+x^{(n)})^2+(\theta_p
y^{(n)}_1+y^{(n)}_2+y^{(n)}_3)^2\\[2mm]
P(y_1,y_2,y_3)&=&3y^2_1+(4\theta^2_p-4\theta_p+3)y_2^2\\
&&+(3\theta^2_p-4\theta_p+4)y^2_3\\
&&+2(2\theta_p+1)y_1y_2+ 2(\theta_p+2)y_1 y_3 \\
&&-2(2\theta^2_p-7\theta_p+2)y_2y_3.
\end{eqnarray*}

The average magnetization $ m $ for the $n$th
generation is then given by
\begin{equation}\label{mag}
m = 2-\frac{4(1+x^{(n)})Y^{(n)}}
{3(1+x^{(n)})^2-2(1+x^{(n)})Y^{(n)}+3(Y^{(n)})^2}.
\end{equation}
where $Y^{(n)}=y^{(n)}_1+y^{(n)}_2 +y^{(n)}_3$. It is quite obvious to note from \eqref{rc} that the set $D=\{(x,0,0,0): x\in R_+\}$ is invariant with
respect to that dynamical system. In this case the system is
reduced to the following one:
\begin{equation}\label{rc1}
f(x)=\frac{1}{2\theta_1}\bigg(\frac{(\theta_p+1)x+2)}{\theta_p+x}\bigg)^2
\end{equation}

Given the conditions $\t_p<1$ and $\t_1>1$ only one fixed
point $x^*$ of $f(x)$ can be found.

%\section{The Phase diagram}

The derived recursion relations \eqref{rc} provide us (numerically) with the
exact phase diagram in the $(T/J_1, -J_p/J_1)$
space. Starting from random initial conditions (subject to the constraint
$y_1,y_2,y_3\neq 0$), we may observe the behavior of the recurrence
relations \eqref{rc} after a large number of numerical iterations. The
phases are characterized by the sequence of stable points of the
recursion relations. Namely, in the simplest case a fixed point
$(x^*,y^*_1,y^*_2,y^*_3)$ is reached. This point corresponds to a
paramagnetic phase if $y^*_1=y^*_2=y^*_3=0$, or to a ferromagnetic
phase if $y^*_1,y^*_2,y^*_3\neq 0$.  The lower-order commensurate
phases (short period) are described by a sequence of a few fixed
points while the higher-order commensurate phases (large period) or
incommensurate phases (infinite period) are described by the
quasicontinuous  or  continuous attractors, respectively. The
distinction between a truly aperiodic case and one with a very
long period is difficult to make numerically.

\begin{figure}
\centering
\includegraphics[width=0.9\columnwidth,clip]{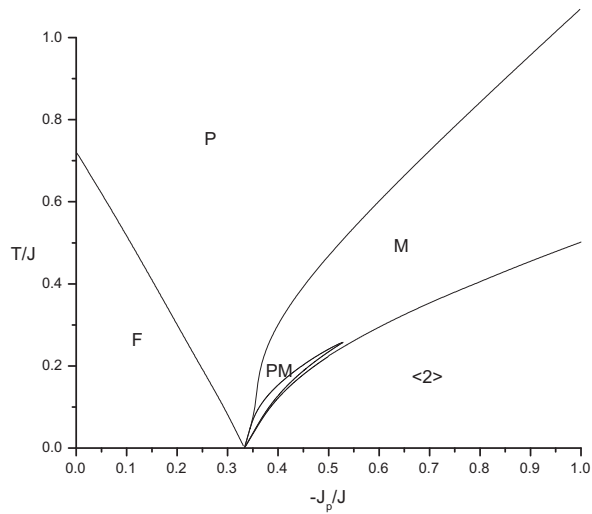} %
\caption{Phase Diagram of the model: $F$ means ferromagnetic, $P$
paramagnetic, $M$ modulated, $PM$ paramodulated, and $<2>$ the
antiphase structure, i.e. the structure with period 4.}\label{PD1}
\end{figure}

The obtained phase diagram is presented in Fig. \ref{PD1}. At
$T=0$ only two different ground-states are encountered: the
ferromagnetic states for $-J_p/J_1<1/3$, and states of period $4$
(for example states with structure $(1122)$,$(1133)$,$(2233)$) for
$-J_p/J_1>1/3$. Such states are called antiphase and denoted  as $<2>$.
The main feature to be noted is the existence of a multiphase at finite
temperature ($T=0, -J_p/J_1=1/3$); this is where the paramagnetic (P),
ferromagnetic (F), antiphase $<2>$ (see below), modulated (M)
(this phase contains both commensurate and incommensurate regions)
and (PM) paramodulated phases, meet. We note the finding of a new phase we refer to as paramodulated, which is characterized by 2-periodic points of the dynamical system \eqref{rc1} and lies inside the modulated phase. From \eqref{mag} one can see that in this region the
average magnetization is the same as in the Paramagnetic one. It is important to note that such a phase is absent
in the case of the Ising model, thus
exhibiting a siginificant difference between phase diagrams for Potts and Ising
models. The rest of the diagram is quite
similar to the one obtained by Vannimenus \cite{V} for the Ising
model with similar interactions.

Below we detail out the critical lines encountered in
the phase diagram.

{\em The Paramagnetic Phase}. The transition lines of the para-ferro and para-modulated are found
to be continuous.  Such lines are obtained by linearizing the system
\eqref{rc} around the fixed point $(x^*,0,0,0)$. Note that the
parameters $y_1$, $y_2$ and $y_3$ vanish in the region $R$.
The variable $x^*$ is unaffected in first order in $y_1$, $y_2$
and $y_3$.

The eigenvalue equation of the linearized system has the following
form:
\begin{eqnarray}\label{egv1}
&&\lambda^3 +\frac{2(x^{*2}-\theta^2_px^*
-2\theta_p)}{(\theta+x^*)[(\theta_p+1)x^*+2]}\lambda^2\nonumber \\
&& +
\frac{4(1-\theta_p)x^*[(\theta_p+1)x^{*2}+(\theta_p+1)^2x^*+2]}{(\theta_p+x^*)[(\theta_p+1)x^*+2]^2}\lambda\nonumber\\
&& +
\frac{8(1-\theta_p)^2(\theta_p+2)x^{*2}}{(\theta_p+x^*)[(\theta_p+1)x^*+2]^2}=0
\end{eqnarray}

The fixed point is linearly stable if the
eigenvalues has moduli smaller than one. It is interesting to note that in the Paramagnetic
case from \eqref{mag}, one may observe the average magnetization value
as being equal to 2.

To find out the transition lines it is necessary to examine several cases
with respect to whether the the eigenvalues are real or complex.

{\em The Para-Ferro Transition.}
When the eigenvalues are real, the transition line will be
characterized by the criterion that the largest (in absolute
value) eigenvalue should be equal to unity. This determines the
stability limit line we were looking for (since all fixed points
are linearly stable if the eigenvalues have moduli smaller than
one).

The condition that the largest eigenvalue is equal to 1 becomes
\begin{eqnarray}\label{PF}
&&(6\theta_p +2)x^{*3}+(3\theta^3_p -12\theta^2_p-7\theta_p
+28)x^{*2} \nonumber\\
&&-(4\theta^2_p +8\theta_p -12)x^* -4\theta_p=0
\end{eqnarray}
with $ x^*$ positive. From this equation one can find $x^*$ and
taking into account that is a fixed point of \eqref{rc1}, one gets
an equation of the para-ferro transition line which will be in the
form $\theta_1=g(\theta_p)$ for some function $g$.

Note that the case $\theta_p=1$ (i.e. $J_p=0$) corresponds to the
simple Potts model with nearest neighbor interactions and one
recovers the well-known result for the critical temperature: $
\exp(J/T_c)=4$ \cite{g, PLM}.

Observations of \eqref{PF} show that at low temperatures
($\theta_p \ll 1$) from \eqref{rc1} one has $x^* \sim {\theta_p
/3}$ and $\theta_1 \sim {(27/4)\theta^{-3}_p}$. In terms of $T$
and $ p\equiv-J_p/J $, the equation of the transition line is
given by
$$ 1-3p =T\log\bigg(\frac{27}{4}\bigg) \qquad (T\rightarrow 0),$$
which is in agreement with the slope obtained numerically (see
Fig. \ref{PD1}).

{\em The Para-Modulated Transition}.
When the three eigenvalues are a pair of complex conjugates and one real, then the
fixed point is approached in an oscillatory way and stability
is achieved if the absolute values of all eigenvalues are less than 1.
The critical transition line will then be characterized
by the criterion that the modules of the complex eigenvalues are
equal to unity. Therefore, the instability occurs when eigenvalue
$ \lambda= i$. Hence, the eigenvalue equation \eqref{egv1} is
reduced to \begin{eqnarray}\label{PM} &&(5\theta^2_p +2\theta_p
-3)x^{*3}+(5\theta^3_p +6\theta^2_p +\theta_p) x^{*2}\nonumber \\
&&+4(\theta^2_p +3\theta_p -1)x^* +4\theta_p=0 .
\end{eqnarray}

If  $ 5\theta^2_p +2\theta_p -3 > 0 $, i.e. $ \theta_p >
\frac{3}{5}$, then equation \eqref{PM} does not allow for any positive
solution. Therefore, the transition  exists only if $ \theta_p <
\frac{3}{5}$, that is $ \frac{J_p}{T}< \log\frac{3}{5}$,
$$ -\frac{J_p}{J}>\frac{T}{J} \log\frac{5}{3}  $$
and it corresponds to the asymptote of the transition line for
large $T$ in Fig. \ref{PD1}

Observations of \eqref{PM} show at low temperatures ($\theta_p \ll
1$), $x^* \sim {\theta_p }$ and from (\ref{egv1})  one finds
$\theta_1 \sim {(1/2)\theta^{-3}_p}$. In terms of $T$ and $ p
\equiv -J_p/J $, the equation of the transition line becomes
$$ 1-3p = {-\frac{T}{J}\log 2}  \qquad (T\rightarrow 0) $$
which is in agreement with the slope obtained numerically.

{\em The Paramodulated Phase}.

\begin{figure}[thbp]
\centering
\includegraphics[width=0.9\columnwidth,clip]{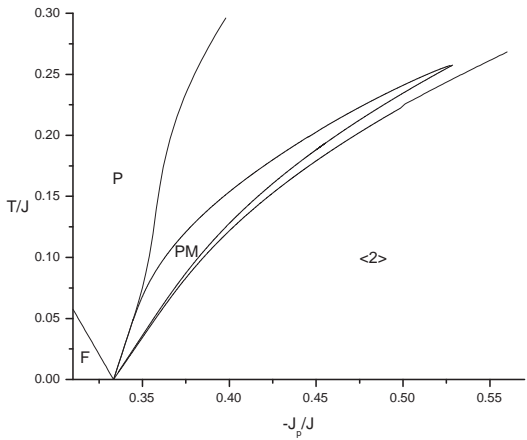}
\caption{Enlargement of Fig. \ref{PD1} around the PM phase}
\label{PM}
\end{figure}

\begin{figure}[htbp]
\centering
\includegraphics[width=0.9\columnwidth,clip]{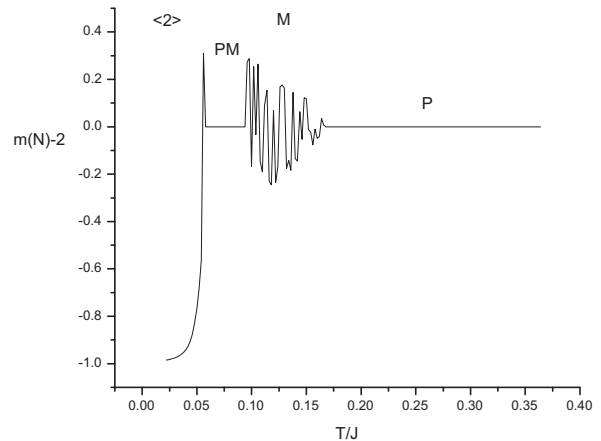}
\caption{The average magnetization versus $T/J$ for
$-J_p/J=0.36$.}\label{m1}
\end{figure}

\begin{figure}[htbp]
\centering
\includegraphics[width=0.9\columnwidth,clip]{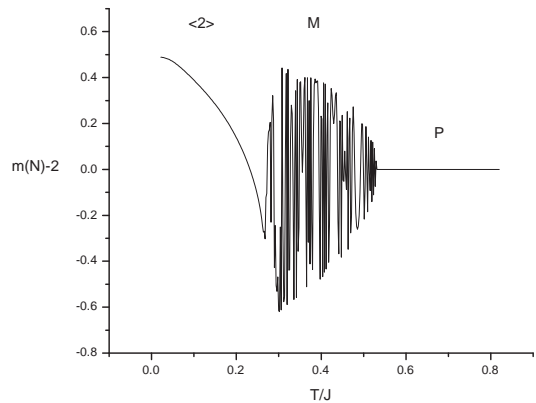}
\caption{The average magnetization versus $T/J$ for
$-J_p/J=0.56$.}\label{m2}
\end{figure}

As noted earlier that the set $D=\{(x,0,0,0): x\in R_+\}$ is invariant
for the dynamical system \eqref{rc}, one can consider
its restriction to this set, which is given by \eqref{rc1}.
Numerical investigations of the dynamics of \eqref{rc1} show that
in some values of the parameters $T/J_1$ and $-J_p/J_1$ it has
2-periodic points (note that no other periodic points are present). In
the plane $(T/J_1,J_p/J_1)$ such values form an island inside the
modulated region (see Fig. \ref{PM}). In this island, the average
magnetization $m$ is equal to 2 as in the paramagnetic case.
Therefore, we denote it as {\it paramodulated} (PM) phase.
A similar phase is not present in the case of the Ising model, hence marking a
difference between Potts and Ising models \cite{V,MTA}. In
Fig. \ref{m1} and Fig. \ref{m2} plots of the values of
magnetization $m$ are given at selected values $-J_p/J=0.36$ and
$-J_p/J=0.56$, respectively. Note that this phase also issues from
the multi-critical point (the Lifshitz point) $T=0, -J_p/J_1=1/3$.

%\section{Discussions}

{\em Discussions}. We have found an exact phase diagram of the Potts model with next nearest neighbor
interactions on the Bethe lattice of order two. The diagram consists of five phases: ferromagnetic, paramagnetic, modulated,  antiphase and paramodulated, all meeting at the Lifshift point i.e. $p=1/3$.
A distinctive feature of the diagram is seen in the existence of a new phase called the paramodulated phase found at low temperatures and characterized by 2-periodic points of the dynamical system \eqref{rc1}. This phase lies inside the modulated phase and is inherent in the Potts model as no analogue can be found within the Ising setting.
We believe the appearance of such a phase to essentially be another form of symmetry resulting in the increasing of the number of spin from two to three.

\acknowledgements This work was partially supported by  SAGA Fund P77c by the Ministry of Science, Technology and Innovation (MOSTI) through the Academy of Sciences Malaysia(ASM) \vspace{-6pt}.


\begin{thebibliography}{1000}

\bibitem{NS} M.P. Nightingale, M. Schick,
{\it J.Phys. A: Math. Gen.} {\bf 15}, L39-L42  (1982).

\bibitem{Ba} R.J. Baxter, {\it Exactly  Solved Models in Statistical Mechanics},
(Academic Press, London/New York, 1982).

\bibitem{DGM} S.N. Dorogovtsev, A.V. Goltsev, J.F.F.Mendes, {\it Eur. Phys. J. B} {\bf 38} 177-182 (2004).

\bibitem{Mo} J.L. Monroe,  {\it
Physics Lett.  A.} {\bf 188},  80-84 (1994); {\it Phys. Rev. E}
{\bf 67} , 017103 (2003).

\bibitem{T} P.N. Timonin,  {\it JETP} {\bf 99} ,
1044--1053 (2004).

\bibitem{Po} R.B. Potts,
{\it  Proc. Cambridge Philos. Soc.} {\bf 48}, 106--109 (1952).

\bibitem{W} F.Y. Wu, {\it Rev. Mod. Phys.}
{\bf 54} , 235--268 (1982).

\bibitem{CMC} S. Coutinho, W. A. M. Morgado, E. M. F. Curado,  L. da
Silva, Phys. Rev. {\bf 74}, 094432 (2006)

\bibitem{PLM} F. Peruggi, F. di Liberto, G. Monroy,
 {\it J. Phys.
A} {\bf 16} , 811--827 (1983);
{\it Physica A } {\bf 141}  151--186 (1987).

\bibitem{g} N.N. Ganikhodjaev, {\it Theor. Math. Phys.} {\bf 85} , 1125--1134 (1990).

\bibitem{Ising} The Ising model with competing interactions was
originally considered by Elliot \cite{E} in order to describe
modulated structures in rare-earth systems. In \cite{BB} the
interest to the model was renewed and studied by means of an
iteration procedure. The Ising type models on the Bethe lattices
with competing interactions appeared in a pioneering work
Vannimenus \cite{V}, in which the physical motivations for the
urgency of the study such models were presented. In \cite{YOS,TY}
the infinite-coordination limit of the model introduced by
Vannimenus was considered. It was also found a phase diagram which
was similar to that model studied in \cite{BB}. In
\cite{MTA},\cite{SC} other generalizations of the model were
studied.

\bibitem{E} R.J. Elliott, {\it Phys. Rev.}, {\bf 124}, 340--345 (1961).

\bibitem{BB} P.Bak, J.von Boehm, {\it Phys. Rev. B},{\bf 21},
5297--5308 (1980).

\bibitem{V} J.Vannimenus,  {\it Z.Phys. B} {\bf
43}, 141--148 (1981).

\bibitem{YOS} C.S.O.Yokoi, M.J. Oliveira, S.R. Salinas, {\it Phys. Rev.
Lett.}, {\bf 54}, 163--166 (1985).

\bibitem{TY} M.H.R. Tragtenberg, C.S.O. Yokoi, {\it Phys. Rev. B}, {\bf
52}, 2187--2197 (1995).

\bibitem{MTA} M. Mariz, C.Tsalis, A.L.Albuquerque, {\it Jour. Stat. Phys.} {\bf
40}, 577--592 (1985).

\bibitem{SC} C.R. da Silca, S. Coutinho, {\it Phys. Rev. B},{\bf 34},
7975-7985 (1986).

\bibitem{KF} F.A. Kassan-Ogly,  B.N. Filippov, Jour. Magnetism. Magn. Mat. {\bf 300},
559--562 (2006); F.A. Kassan-Ogly, I.V. Sagaradze, Phys. Metals
Metallography {\bf 100} 201-207 (2005).

\bibitem{PT} I. Peschel, T. T. Truong, Jour. Stat. Phys. {\bf 45}
1572--9613 (1986).

\bibitem{Ma} M.C. Marques, {\it J.Phys. A: Math. Gen.} {\bf 21},
1061-1068 (1988).

\bibitem{Ta} M. Tarnawski J. Phys.: Condens. Matter {\bf 1}, 1849-1854 (1989); {\bf 2} 8599--8613 (1990).

\bibitem{BW} J.R. Banavar, F.Y Wu, Physical Review B {\bf 29}
1511-1513 (1984)

\bibitem{I} M. Itakura, Phys. Rev. B {\bf 55}, 48 - 51 (1997).

\bibitem{one} One dimensional Potts models with next nearest
neighbor interactions were studied in \cite{KF},\cite{CS}. Recently, in \cite{gmm} the Potts model with
one level competing intercations on a Bethe lattice has been exactly solved.

\bibitem{CS} S.-C. Chang, R. Shrock,     Inter. J. Mod. Phys. B {\bf 15}
443-478 (2001). R. Shrock, S.-H. Tsai, Phys. Rev. E {\bf 55}
5184-5193 (1997).


\bibitem{gmm} N.N. Ganikhodjaev, F.M. Mukhamedov, J.F.F. Mendes, {\it Jour. Stat. Mech.}  P08012 (2006).



\bibitem{Au} S. Aubry, in Solutions and Condesend Matter, edited by A.R. Bishop and T.Schneider (Springer-Verlag,
Berlin, 1978).


\end{thebibliography}
\end{document}